\documentclass[12pt,onecolumn]{article}

\usepackage{enumerate}

\usepackage{hyperref}
\usepackage{graphicx}
\usepackage{amsmath}
\usepackage{amssymb}
\usepackage{amsthm}
\usepackage{epsfig}
\usepackage{subfigure}

\newcommand{\be}{\begin{equation}}
\newcommand{\ee}{\end{equation}}
\newcommand{\ba}{\begin{array}}
\newcommand{\ea}{\end{array}}
\newcommand{\bea}{\begin{eqnarray}}
\newcommand{\eea}{\end{eqnarray}}

\newcommand*{\cL}{\mathcal{L}}
\newcommand*{\cK}{\mathcal{K}}

\newcommand{\calL}{{\cal L }}

\newcommand{\nn}{\nonumber}

\newcommand{\trace}{\mathop{\mathrm{Tr}}\nolimits}

\newcommand{\tr}{\trace}

\newcommand{\pf}{\mathop{\mathrm{Pf}}\nolimits}

\newcommand*{\ket}[1]{|#1\rangle}

\newcommand*{\kket}[1]{|#1\rangle\hspace{-0.55ex}\rangle}
\newcommand*{\sspr}[2]{\langle\hspace{-0.55ex}\langle #1|#2\rangle\hspace{-0.55ex}\rangle}

\newtheorem{lemma}{Lemma}

\newtheorem{theorem}{Theorem}

\newcommand*{\cS}{\mathcal{S}}

\title{Classical simulation of\\
 dissipative fermionic linear optics}
\author{Sergey Bravyi and Robert K\"onig\\
\small IBM T.J. Watson Research Center, Yorktown Heights, NY 10598, USA}

\begin{document}
\maketitle
\begin{abstract}
Fermionic linear optics is a limited form of quantum computation
which is known to be efficiently simulable on a classical computer. We revisit and extend this result by enlarging the set of available computational gates:
in addition to  unitaries and measurements, we allow  dissipative  
evolution governed by a Markovian master equation with linear Lindblad operators. We show that this more general form of fermionic computation is also simulable efficiently by classical means. Given a system of $N$~fermionic modes, our algorithm simulates any such gate in time $O(N^3)$
while a single-mode measurement is simulated in time $O(N^2)$.
The steady state of the Lindblad equation can be computed in time $O(N^3)$.
 \end{abstract}

\section{Introduction}
Quantum dynamical processes have inspired numerous models of computation: adiabatic quantum computation~\cite{Farhi01}, dissipative quantum computation~\cite{verstraeteetal09,Klieschetal11}  or computations based on modular functors~\cite{freedman02} are all examples of computational models associated with certain time-evolving quantum systems. In trying to characterize their computational power, arguably the most practically relevant question is how they compare to universal classical respectively quantum computation.  Indeed, if computations in a model are efficiently simulable on a classical computer, the corresponding physical system may be accessible to numerical studies, but  is unlikely to be a suitable substrate for building a quantum computer.  In contrast, simulability by quantum circuits means that the underlying physics could be studied using a quantum computer, the prime application of such machines originally envisioned by Feynman~\cite{feyman82}. Finally, in cases where the computational resources provided by a model are sufficient to implement universal quantum computation,  the corresponding physical system is a candidate for the realization of a quantum computer.

A number of physically motivated models can be understood as the result of restricting the available set of initial states, gates and measurements in the standard quantum circuit model. For  topological~\cite{Kitaev03,preskill,Freedmanetal03} or permutational~\cite{Jordan10} quantum computing, there is a preferred initial (vacuum) state, the available gates represent braid group generators or transpositions and there is a set of allowed (charge) measurements. In bosonic linear quantum optics~\cite{KnillLaflammeMilburn01}, the available repertoire includes preparation of the vacuum initial state, single photon sources, beam splitters, phase shifters and photo-detectors. In fermionic quantum optics~\cite{TerhalDiVincenzo02,Knill01}, we permit preparation of the vacuum state, free unitary evolution and occupation number measurements.  The computational power of topological computing depends on the representation: for example, it is universal in the case of the Fibonacci model but classically simulable for Ising anyons. Bosonic linear optics was shown to be universal~\cite{KnillLaflammeMilburn01} for quantum computation. Fermionic linear optics can be simulated efficiently on a classical computer~\cite{TerhalDiVincenzo02,Knill01} (see below).

The model of dissipative quantum computing is of a conceptually different origin: it is the result of considering more general quantum dynamics beyond unitary evolutions. Here the standard unitary gate set is augmented or completely replaced by `dissipative gates'. Mathematically, the latter are completely positive trace-preserving maps corresponding to the time-evolution under a Markovian master equation. Such Markovian evolution is usually associated with noise in implementing a quantum computer, and one seeks to reduce its detrimental effect by the use of error-correction. Recently, however, it was realized that purely dissipative processes can actually be useful for quantum computation (see e.g.,~\cite{Griesseretal05,verstraeteetal09,Weimer10}). In~\cite{verstraeteetal09}, it was shown that dissipative quantum computation is universal for quantum computation. Conversely, Markovian dynamics with local Lindblad operators can be simulated efficiently (i.e., with polynomial overhead) on a quantum computer~\cite{Klieschetal11}.

Here we consider a model of computation that extends fermionic linear optics with dissipative processes. We will show that this extended model is still efficiently simulable classically. Since our simulation algorithm does not rely on Trotter-type expansions (as e.g.,~\cite{Klieschetal11}), dissipative processes can be simulated exactly for any evolution time without affecting complexity or accuracy. The algorithm reproduces the statistics of measurement outcomes and also provides a complete description of the state at any instant in the computation. Additionally, the Liouvillian may be non-local, but we restrict to Lindblad operators which are linear in fermionic annihilation and creation operators.

The physics underlying fermionic linear optics --   non-interacting fermions -- encompasses a number of systems of interest in condensed matter physics, including  Kitaev's Majorana chain~\cite{Kitaev00}  or honeycomb model~\cite{Kitaev05}. Such systems exhibit topological order and could be used as fault-tolerant  quantum memories or topological quantum computers. The classical simulation algorithm discussed here may be particularly suited to assess their potential to serve such information-processing purposes. Typically, this involves studying not only the dynamics of the system, but also the effect of simple manipulations (such as syndrome measurements or error correction). As an example, the known simulation technique in the non-dissipative case has been used to study the beneficial effect of disorder on the performance of the Majorana chain as a quantum memory~\cite{BraKoe11}. The extended toolkit provided here may be applied to evaluate the robustness of certain proposals, e.g., for state transfer~\cite{Yaoetal11,Yaoetal11b}, in the presence of dissipation.

\section{Fermionic linear optics\label{sec:flo}}
Fermionic linear optics is defined in terms of
$N$ creation and annihilation operators~$a_j^\dagger$ and~$a_j$ satisfying Fermi-Dirac canonical commutation relations~$\{a_j,a_k\}=0$ and~$\{a_j,a_k^\dagger\}=\delta_{j,k}I$. We will refer to such a family of operators $\{a_j,a_j^\dagger\}_{j=1}^N$ as $N$~Dirac fermions or  modes to distinguish them from Majorana fermions introduced below (the latter will be denoted by the letter~$c$ instead of~$a$).  The corresponding Hilbert space is spanned by the number states
\begin{align}
\ket{n_1,\ldots,n_N}=(a_1^\dagger)^{n_1}\cdots (a_N^\dagger)^{n_N}\ket{0}\qquad\textrm{ where }n_j\in \{0,1\}\ ,\label{eq:numberstate}
\end{align}
and $\ket{0}$ is the fermionic vacuum state satisfying $a_j\ket{0}=0$ for all~$j$. A state of the form~\eqref{eq:numberstate} is an eigenstate of the occupation number operator $N_j=a_j^\dagger a_j$ with eigenvalue~$n_j$.

The allowed operations defining fermionic linear optics are:
\begin{enumerate}[(i)]
\item
preparation of the fermionic vacuum~$\ket{0}$\label{it:vacuumstateprepare}
\item
measurement of the occupation numbers $N_j$, $j\in\cS$ for any subset $\cS$ of modes,\label{it:occupationnumber}
\item\label{it:unitaryevolve}
evolution under a quadratic fermion Hamiltonian $H$ for a time~$t$, that is, according to the equation of motion
\begin{align}
\frac{d}{dt}\rho=-i [H,\rho]\ .\label{eq:closed}
\end{align} Fermionic parity preservation implies that~$H$ is a linear combination of terms
 of the form~$\epsilon_j a_j^\dagger a_j$ giving energy~$\epsilon_j\in\mathbb{R}$ to mode~$j$, `hopping terms' of the form~$t_{j,k}a_j^\dagger a_k+t^*_{j,k}a_k^\dagger a_j$ where $t_{j,k}\in\mathbb{C}$ for $j\neq k$ and  `pair creation/annihilation terms' of the form $s_{j,k}a_j^\dagger a_k^\dagger+s^*_{j,k}a_ka_j$, $s_{j,k}\in\mathbb{C}$.
\end{enumerate}
These operations can be performed in an arbitrary order, and, in particular, may depend on  measurement results obtained in the course of the computation. The computation concludes with a final measurement whose outcome is the classical result produced by the computation.

Simulating such a fermionic linear optics computation on a classical computer amounts to sampling from the distribution of measurement outcomes at the end of the computation. A classical polynomial-time (in $N$) algorithm for doing so was found by Terhal and DiVincenzo~\cite{TerhalDiVincenzo02} and Knill~\cite{Knill01}. A stronger form of simulation outputs a description of the state at any instant during the computation. An efficient algorithm for this problem was provided by Bravyi~\cite{B04}. We review these techniques in Section~\ref{sec:classicalsimulation}.

Fermionic linear optics was motivated by the quantum universality of bosonic linear optics~\cite{KnillLaflammeMilburn01}, but is closely related to another computational model: the unitaries~\eqref{it:unitaryevolve} are so-called matchgates. The corresponding computational model -- matchgate computation --  was earlier shown to be classically simulable by Valiant~\cite{Valiant01}. Josza and Miyake~\cite{Jozsamiyake08} extended the simulation results by showing that more general initial (product) states can be allowed without losing classical simulability. They also showed that a slight modification of Valiant's gate set provides quantum universality (see also~\cite{beenakker04} for a physically motivated universal gate set extending fermionic linear optics).
 More recently, new and elegant characterizations of the power of fermionic linear optics have been provided: Jozsa and coauthors~\cite{Jozsaetal09} showed that the model is equivalent to space-bounded quantum computation (see also~\cite{Bravyikitaev02}), and van den Nest~\cite{denNest11} gave a characterization in terms of linear threshold gates.

\section{Dissipative fermionic linear optics\label{sec:dissipativeqcomp}}

Here we generalize these results as follows: in addition to the unitary set of gates defined by~\eqref{it:unitaryevolve}, we allow dissipative processes. More specifically, we replace~\eqref{it:unitaryevolve} with
\begin{enumerate}[(i')]\setcounter{enumi}{2}
\item\label{it:lindbladevolve}
 evolution under the Lindblad master equation
\begin{align}
\frac{d}{dt}\rho=\cL(\rho):=-i [H,\rho]+\sum_\mu \left(2L_\mu\rho L_\mu^\dagger-\{L_\mu^\dagger L_\mu,\rho\}\right)\ ,\label{eq:lindbladevolution}
\end{align}
for a time~$t$, where $H$ is a quadratic fermion Hamiltonian and each Lindblad operator $L_\mu$ is a linear combination $L_\mu=\sum_j \alpha_{\mu,j}a_j^\dagger+\beta_{\mu,j}a_j$ with $\alpha_{\mu,j},\beta_{\mu,j}\in\mathbb{C}$.
\end{enumerate}
We show that a fermionic computation composed of~\eqref{it:vacuumstateprepare},\eqref{it:occupationnumber} and~\eqref{it:lindbladevolve} can be efficiently simulated on a classical computer.

\subsubsection*{Outline}
The remainder of this paper is structured as follows.  In Section~\ref{sec:background}, we discuss  fermionic Hamiltonians and Liouvillians  and we show that dissipative dynamics with linear Lindblad operators preserves the set of Gaussian fermionic states. In Section~\ref{sec:classicalsimulation}, we review the classical simulation of fermionic linear optics.
In Section~\ref{sec:solution} we explain how to simulate dissipative fermionic linear optics.

\section{Background\label{sec:background}}
In  this section, we give some background on quadratic open fermion systems with linear Lindblad operators. It will be convenient to work with the Hermitian operators
\begin{align*}
c_{2j-1}=a_j+a_j^\dagger\qquad\textrm{ and }\qquad c_{2j}=i(a_j-a_j^\dagger)\qquad\textrm{ for }j=1,\ldots,N\  .
\end{align*}
The $2N$~operators $c_1,\ldots,c_{2N}$ satisfy the canonical anticommutation relations
\begin{align*}
\{c_j,c_k\}=2\delta_{j,k}I\qquad\textrm{ for all }\qquad j,k=1,\ldots,2N\ ,
\end{align*}
and will be referred to as Majorana fermions or modes. The Hamiltonian and Lindblad operators~\eqref{eq:lindbladevolution} then take the form
\begin{align}
H({\bf H})=\frac{i}{4}\sum_{j,k=1}^{2N}{\bf H}_{jk}c_jc_k=\frac{i}{4}\underline{c}\cdot {\bf H}\underline{c}\qquad L_\mu=\sum_{j=1}^{2N}\ell_{\mu,j}c_j={\underline{\ell}}_\mu\cdot \underline{c}\label{eq:Handlindblad}
\end{align}
where $\underline{c}=(c_1,\ldots,c_{2N})$.  Here ${\bf H}=-{\bf H}^T\in\mathfrak{so}(2N,\mathbb{R})$ is an antisymmetric matrix with real entries. In contrast, the vectors $\underline{\ell}_\mu=(\ell_{\mu,1},\ldots,\ell_{\mu,2N})\in \mathbb{C}^{2N}$ are generally complex-valued.

\subsection{Third quantization description of Liouvillians\label{sec:thirdquantization}}
We will present two methods for simulating dissipative dynamics of the form~\eqref{eq:lindbladevolution}. One of these methods is based on the language of `third quantization'. This refers to the fact that the Liouvillian~$\cL$ is quadratic in a set of fermionic superoperators as discussed by Prosen~\cite{prosen}
 (see also~\cite{ProzenZunkovic10} and Dzhioev and Kosov~\cite{dzhioev11}).  Here we follow his presentation. 

Let~$\cK$ be the $4^N$-dimensional vector space spanned by the monomials~$c^{\underline{\alpha}}=c_1^{\alpha_1}\cdots c_{2N}^{\alpha_{2N}}$ where $\underline{\alpha}=(\alpha_1,\ldots,\alpha_{2N})\in\{0,1\}^{2N}$. To emphasize the vector space structure,  we will write an operator $\rho=\sum_{\underline{\alpha}}\lambda_{\underline{\alpha}}c^{\underline{\alpha}}$ (with $\lambda_{\underline{\alpha}}\in \mathbb{R})$ as $\kket{\rho}=\sum_{\underline{\alpha}}\lambda_{\underline{\alpha}}\kket{c^{\underline{\alpha}}}$ when it is considered as an element of~$\cK$. The space $\cK$ is equipped with the Hilbert-Schmidt inner product $\sspr{\rho}{\sigma}=2^{-N}\tr\rho^\dagger\sigma$, and with respect to the latter, the monomials $\kket{c^{\underline{\alpha}}}$ are an orthonormal basis. The space~$\cK$ decomposes into a direct sum $\cK=\cK^+\oplus \cK^-$ of spaces spanned by monomials~$\kket{c^{\underline{\alpha}}}$ with even $(\cK^+)$ and odd $(\cK^-)$ parity~$\sum_{j=1}^{2N}\alpha_j \mod 2$. In the following, we will often restrict our attention to the subspace~$\cK^+$, although identical arguments apply to~$\cK^-$. By fermionic parity superselection, this is sufficient to cover physical states.

We define~$2N$ pairs $(\hat{a}_{j},\hat{a}_{j}^\dagger)$ of mutually adjoint linear operators acting on~$\cK$ by
\begin{align}
\hat{a}_j\kket{c^{\underline{\alpha}}}=\delta_{\alpha_j,1}\kket{c_j c^{\underline{\alpha}}}\qquad\textrm{ and }\qquad
\hat{a}_j^\dagger\kket{c^{\underline{\alpha}}}=\delta_{\alpha_j,0}\kket{c_j c^{\underline{\alpha}}}\qquad\textrm{ for }j=1,\ldots, 2N  .\label{eq:thirdquantizeddirac}
\end{align}
These satisfy canonical Dirac anticommutation relations
\begin{align*}
\{\hat{a}_j,\hat{a}_k\}=0\qquad\{\hat{a}_j,\hat{a}_k^\dagger\}=\delta_{j,k}\qquad\textrm{ for }j,k=1,\ldots,2N\ ,
\end{align*}
which has motivated the expression `third quantization'. Observe that the state~$\kket{c^{\underline{0}}}=\kket{I}$ corresponding to the identity operator~$I$ is the vacuum state associated with this set of operators, i.e., $\hat{a}_j\kket{I}=0$ for all~$j$.

It is again often more convenient to work with Hermitian Majorana fermions
\begin{align}
\hat{c}_{2j-1}=\hat{a}_j+\hat{a}_j^\dagger\qquad\textrm{ and }\qquad \hat{c}_{2j}=i(\hat{a}_j-\hat{a}_j^\dagger)\qquad\textrm{ for }j=1,\ldots,2N\ .\label{eq:thirdquantizedmajorana}
\end{align}
The restriction~$\cL_+=\cL|_{\cK^+}$ of the Liouvillian~$\cL$ to operators supported on~$\cK^+$ will be called the even part of~$\cL$.
Remarkably, this superoperator  can be expressed as a quadratic form of the operators~\eqref{eq:thirdquantizeddirac} (or equivalently~\eqref{eq:thirdquantizedmajorana}). More precisely, we have
\begin{align}
\kket{\cL(\rho)}&=\widehat{\cL_+}\kket{\rho}\qquad\textrm{ for all }\rho\textrm{ with } supp(\rho)\subset\cK^+\ ,\textrm{ where }\qquad \widehat{\cL_+}=\frac{1}{4}\underline{\hat{c}}\cdot {\bf L}\underline{\hat{c}}\ ,\label{eq:liouvillethird}
\end{align}
and where $\underline{\hat{c}}=(\hat{c}_1,\ldots,\hat{c}_{4N})$ for an antisymmetric complex-valued matrix~${\bf L}=-{\bf L}^T\in\mathfrak{so}(4N,\mathbb{C})$. The matrix~${\bf L}$  depends linearly on ${\bf H}$ and quadratically on the entries of the vectors~$\underline{\ell}_\mu$ specifying the Lindblad operators. Explicitly, it is given by~\cite{prosen}
\begin{align}
\begin{matrix}
{\bf L}_{2j-1,2k-1}&=&{\bf H}_{j,k}-2{\bf M}_{j,k}+2{\bf M}_{k,j}\\
{\bf L}_{2j,2k}&=&{\bf H}_{j,k}+2{\bf M}_{j,k}-2{\bf M}_{k,j}
\end{matrix}\qquad
\begin{matrix}
{\bf L}_{2j-1,2k}&=&4i{\bf M}_{k,j}\\
{\bf L}_{2j,2k-1}&=&-4i{\bf M}_{j,k}
\end{matrix}\label{eq:lmatrix}
\end{align}
with  ${\bf M}_{j,k}=\sum_{\mu}\ell_{\mu,j}\ell^*_{\mu,k}$.
In~\cite{prosen,ProzenZunkovic10}, it is assumed
that the Liouvillian is generic in the sense that ${\bf L}$ is diagonalizable. Here we do not require such an assumption. Note also that in~\cite{Prosenspectr10}, a normal form for such matrices ${\bf L}$ is derived.

Observe that for any integer~$n$ and any state~$\rho$ supported on~$\cK^+$, we have
\begin{align*}
\kket{\cL_+^n(\rho)}=\kket{\underbrace{\cL_+\circ\cdots\circ\cL_+}_n(\rho)}=\widehat{\cL_+}^n\kket{\rho}\ ,
\end{align*}
which implies that the superoperator~$\exp(t\cL_+)$ corresponding to time evolution for a time~$t$ under~\eqref{eq:lindbladevolution} is given by
\begin{align}
\widehat{\exp(t\cL_+)}&=\exp(t\widehat{\cL_+})\ \label{eq:supotimeevolve}
\end{align}
when restricted to operators supported on~$\cK^+$.  Because $\det \exp(A)=\exp(\tr(A))$, the operator~\eqref{eq:supotimeevolve} is invertible with inverse $\exp(-t\widehat{\cL_+})$. 
 
Finally, consider the adjoint Liouvillian defined by
\begin{align}
\cL^\dagger(O)&=i [H,O]+\sum_\mu \left(2L_\mu^\dagger O L_\mu-\{L_\mu^\dagger L_\mu,O\}\right)\ . \label{eq:adjointliouvillian}
\end{align}
Evolution for a time~$t$ under $\cL^\dagger$
generates the adjoint superoperator of~$\exp(t\cL)$, that is, $\tr(\exp(t\cL)(\rho)O)=\tr(\rho\exp(t\cL^\dagger)(O))$ for all operators~$\rho$, $O$.  This follows by comparing the derivative
\begin{align*}
\frac{d}{dt}\tr(O e^{t\cL}(\rho))&=\tr(O\cL(\rho))=\tr(\cL^\dagger(O)\rho)=\frac{d}{dt}\tr(e^{t\cL^\dagger}(O)\rho)\
\end{align*}
and observing that the two expressions agree for~$t=0$. Since~$\exp(t\cL)$ is a completely positive trace-preserving map, the map~$\exp(t\cL^\dagger)$ is unital, that is, $\exp(t\cL^\dagger)(I)=I$. The even part~$\cL^\dagger_+$ of the adjoint Liouvillian takes the form
\begin{align}
\widehat{\cL_+^\dagger}&=\frac{1}{4}\underline{\hat{c}}\cdot {\bf L^\dagger} \underline{\hat{c}}\ \label{eq:daggerform}
\end{align}
where the matrix  ${\bf L}^\dagger=-{\bf L}^*\in \mathfrak{so}(4N,\mathbb{C})$ is the Hermitian conjugate of~${\bf L}$.

\subsection{Gaussian states\label{sec:gaussian}}
 Here we collect a few facts about Gaussian states
 of $2N$~Majorana fermions~$c_1,\ldots,c_{2N}$  (see~\cite{B04} for details and proofs). A Gaussian state~$\rho$ is completely determined by its covariance matrix
\begin{align}
M_{j,k}&=\frac{i}{2}\tr(\rho [c_j,c_k])\qquad\textrm{ for }j,k=1,\ldots,2N\ \label{eq:covariancematrixdef}
\end{align}
and Wick's formula (see e.g.,~\cite[Eq.~17]{B04})
\begin{align}
i^p \tr(\rho c_{j_1}c_{j_2}\cdots c_{j_{2p}})&=\pf(M\left[j_1,\ldots,j_{2p}\right])\qquad\textrm{ for all }1\leq j_1<j_2<\cdots< j_{2p}\leq 2N\ .\label{eq:wick}
\end{align}
Here $M[j_1,\ldots,j_{2p}]$ is the submatrix of~$M$ of size~$2p\times 2p$ obtained by removing all columns and rows except those indexed by $j_1,\ldots,j_{2p}$, and $\pf$~denotes the Pfaffian.  The antisymmetric matrix $M\in\mathfrak{so}(2N,\mathbb{R})$ can be brought into block-diagonal form by a special orthogonal matrix
\begin{align*}
M&=R\bigoplus_{j=1}^N \left(\begin{matrix} 0 & \lambda_j\\
-\lambda_j & 0
\end{matrix}\right)R^T\qquad R\in SO(2N), \lambda_j\in\mathbb{R}\ .
\end{align*}
This is called the Williamson normal form of $M$. The transformation $\underline{c}\rightarrow \underline{c}'=R\underline{c}$, that is,
\begin{align*}
c_{a}'&=\sum_{b=1}^{2N} R_{a,b}c_b\qquad\textrm{ for }a=1,\ldots,2N
\end{align*}
corresponds to the adjoint action of a unitary as explained below. In particular, $c_{1}',\ldots,c_{2N}'$ again satisfy canonical commutation relations. We call these the eigenmodes of~$\rho$. Expressed in terms of these operators,~$\rho$ takes the simple form
\begin{align}
\rho=\frac{1}{2^N}\prod_{j=1}^N (I+i\lambda_j c'_{2j-1}c'_{2j})\ .\label{eq:rhoeigenform}
\end{align}

\subsection{Time evolution of Gaussian states}
\label{sec:gausspres}
Evolution of Gaussian states under the dissipative dynamics described by Eq.~(\ref{eq:lindbladevolution})
is particularly simple due to the following fact.
\begin{lemma}\label{lem:gaussianitypreservation}
Let $\cL$ be a Liouvillian for $N$~fermions
whose unitary part is given by a quadratic Hamiltonian and whose  Lindblad operators are linear in the creation- and annihilation operators (cf.~\eqref{eq:lindbladevolution}). Let $\rho$ be a Gaussian state. Then the time-evolved state~$e^{t\cL}(\rho)$ is Gaussian for all~$t\geq 0$.
\end{lemma}

\begin{proof}
We will construct a family of trace preserving completely positive maps  $\{\Phi_\epsilon\}$
depending smoothly on a parameter $\epsilon\ge 0$
such that each map $\Phi_\epsilon$ preserves the set of Gaussian states and
\be
\label{deriv}
\left. \frac{d}{d\epsilon} \Phi_\epsilon(\rho) \right|_{\epsilon=0} = \cL(\rho)
\ee
for any state $\rho$. This implies that
\[
\rho(t)=\lim_{n\to \infty} \Phi_{t/n}^n(\rho)
\]
is a limiting point of a sequence of Gaussian states. Since the set of Gaussian states is compact,
$\rho(t)$ must be a Gaussian state itself.

Let us first consider a Liouvillian with a single Lindblad operator,
\[
\calL(\rho)=2L\rho L^\dag -\{ L^\dag L , \rho\},
\]
where $L$ is a linear combination of the Majorana operators $c_1,\ldots,c_{2N}$  with complex coefficients.
Decompose $L=K+iM$, where
$K,M$ are real linear combinations of the Majorana operators. In particular, both $K$ and $M$ are Hermitian.
Introduce one ancillary pair of Majorana operators $c_{2N+1}\equiv b_1$  and $c_{2N+2}\equiv b_2$
representing an environment and consider a unitary operator
\[
U_\epsilon=\exp{\left[ -\sqrt{2\epsilon}(K b_1 + M b_2)\right]}\ .
\]
Simple algebra shows that
\bea
U_\epsilon \eta U_\epsilon^\dag &=& \eta -\sqrt{2\epsilon} [Kb_1+Mb_2,\eta] -2\epsilon(Kb_1+Mb_2)\eta (Kb_1+Mb_2) \nn \\
&& +\epsilon\{ (Kb_1+Mb_2)^2,\eta\} + O(\epsilon^{3/2}) \label{UetaU}
\eea
for any state $\eta$. Define a map
\be
\label{Phi}
\Phi_{\epsilon}(\rho)=\mathrm{Tr}_E \, U_\epsilon\,  \rho \rho_E U_\epsilon^\dag\ .
\ee
Here $\mathrm{Tr}_E$ represents the partial trace over the environment, and $\rho_E$ is the initial state of the environment
which we choose as the vacuum state, that is,
\[
\rho_E=\frac12(I-ib_1 b_2)\ .
\]
Note that the terms in Eq.~(\ref{UetaU}) that contain half-integer powers of $\epsilon$ do not contribute to $\Phi_{\epsilon}(\rho)$
since $\mathrm{Tr}_E \, b_1 =\mathrm{Tr}_E \, b_2=0$.
Using the identities
\[
\mathrm{Tr}_E\,  b_1 b_2 \rho_E  = i \quad \mbox{and} \quad \mathrm{Tr}_E\,  b_1 \rho_E b_2 = -i
\]
it is easy to check that
\be
\label{deriv1}
\Phi_{\epsilon}(\rho)=\rho+ \epsilon \calL(\rho) + O(\epsilon^{2})\ .
\ee
Given a general Liouvillian with multiple Lindblad operators and the unitary evolution term as in Eq.~\eqref{eq:lindbladevolution}, the desired  map $\Phi_{\epsilon}$ can be constructed by taking a composition
\[
\Phi_{\epsilon}(\rho)=e^{-iH\epsilon} \Phi_{\epsilon}^{(1)} \circ \ldots \circ \Phi_{\epsilon}^{(m)} (\rho) e^{iH\epsilon},
\]
where $\Phi_{\epsilon}^{(\mu)}$ is the map defined above with $L=L_\mu$.
By construction, $\Phi_{\epsilon}$ is a finite composition of unitary evolutions under quadratic Hamiltonians,
addition of ancillary two-mode vacuum states $\rho_E$, and partial traces over some pairs of modes.
It is well-known that all these operations preserve the set of Gaussian states, see e.g.~\cite{B04}.
Finally, Eq.~(\ref{deriv}) follows from Eq.~(\ref{deriv1}) and the product rule for derivatives.
\end{proof}

\section{Classical simulation of fermionic linear optics\label{sec:classicalsimulation}}
In this section, we review the known simulation techniques
 for a computation composed of the operations~$(i)$--$(iii)$ introduced in Section~\ref{sec:flo}.  The basis of the simulation algorithm is the fact that the vacuum state~$\ket{0}$ prepared by~$(i)$ is a Gaussian state and all subsequent operations preserve the Gaussian nature of the state. 

The covariance matrix~$M(0)$ of the vacuum state~$\ket{0}$ at the beginning of the computation is given by the non-zero entries
\begin{align}
M(0)_{2j-1,2j}&=1 \qquad\textrm{ for } j=1,\ldots,2N\
\end{align}
above the main diagonal. The remaining task is to find update rules for the covariance matrix.  These updates can be done efficiently as discussed below: measuring (time) complexity in terms of the number of additions, multiplications, and divisions on complex numbers that are required,  measurements and unitary  gates can be simulated in time~$O(N^3)$.

\subsection{Simulating measurements\label{sec:measurement}}
Here we describe the method from~\cite{BraKoe11} which is a more efficient version of Terhal and DiVincenzo's algorithm~\cite{TerhalDiVincenzo02} (the latter has time complexity $O(N^4)$ when measuring $N$~modes).

Consider a (non-destructive) measurement of the occupation number $N_j=a_j^\dagger a_j=\frac{1}{2}(I-ic_{2j-1}c_{2j})$ in a Gaussian state~$\rho$ with covariance matrix~$M$.
By definition~\eqref{eq:covariancematrixdef}, the probability of obtaining outcome~$1$ when measuring~$\rho$ is
\begin{align}
P_j(1)&=\tr(N_j\rho)=\frac{1}{2}(1-M_{2j-1,2j})\ .\label{eq:measurementprob}
\end{align}
Let $\Pi_j(n_j)=(a_j^\dagger a_j)^{n_j} (a_ja_j^\dagger)^{1-n_j}$ be the projection corresponding to the measurement outcome $n_j\in\{0,1\}$. The post-measurement state
\begin{align}
\rho(n_j)&=\frac{\Pi_j(n_j)\rho \Pi_j(n_j)}{P_j(n_j)} \label{eq:expr}
\end{align}
is Gaussian, as shown in~\cite{B04}. Its covariance matrix~$M^{(j)}(n_j)$ can be computed from~\eqref{eq:measurementprob} and~\eqref{eq:expr} using Wick's theorem~\eqref{eq:wick}, giving (cf.~\cite[Eq.~(7.3)]{BraKoe11})
\begin{align}
M^{(j)}(n_j)_{p,q}=M_{p,q}-\frac{(-1)^{n_j}}{2P_j(n_j)}M_{2j-1,p}M_{2j,p}+\frac{(-1)^{n_j}}{2P_j(n_j)}M_{2j-1,q}M_{2j,p}\ .\label{eq:covariancematrixupdaterule}
\end{align}
The following algorithm then simulates  a measurement of~$N_j$: first, compute~\eqref{eq:measurementprob} and sample a bit $n_j\in\{0,1\}$~according to the probability distribution $\Pr[n_j=1]=P_j(1)$. Then update the covariance matrix according to~\eqref{eq:covariancematrixupdaterule}.  This requires~$O(N^2)$ computational steps and one (non-uniform) bit of randomness.

A measurement of a subset $\cS=\{j_1,\ldots,j_{|\cS|}\}$ of modes can be simulated by iterative use of this procedure, using the recursion relation
\begin{align*}
P^{j_1\ldots j_{\ell-1}}_{j_\ell}(n_{j_\ell}|n_{j_1}\cdots n_{j_{\ell-1}})&=\tr(\Pi_{j_\ell}(n_{j_\ell})\rho(n_{j_1}\cdots n_{j_{\ell-1}}))\\
\rho(n_{j_1},\ldots,n_{j_\ell})&=\frac{\Pi_{j_\ell}(n_{j_\ell})\rho(n_{j_1},\ldots,n_{j_{\ell-1}})
\Pi_{j_{\ell}}(n_{j_\ell})}{P^{j_1\ldots j_{\ell-1}}_{j_\ell}(n_{j_\ell}|n_{j_1}\cdots n_{j_{\ell-1}})}
\end{align*}
for the probability and post-measurement state  after the measurement of the $j_{\ell}$-th mode, given that the measurements of $N_{j_1},\ldots,N_{j_{\ell-1}}$ resulted in  the sequence~$n_{j_1},\ldots,n_{j_{\ell-1}}\in\{0,1\}$. This can be done in time~$O(|\cS|\cdot N^2)$ using $|\cS|$~random bits.

\subsection{Simulating unitary evolution\label{sec:unitaryevolve}}
The classical simulation of unitary dynamics (see~\cite{Knill01,TerhalDiVincenzo02,Jozsamiyake08}) is particularly instructive for our generalization to the dissipative case. Consider time evolution under a quadratic Hamiltonian
$H=H({\bf H})$ (cf.~\eqref{eq:Handlindblad}) for some time~$t$, starting from a Gaussian initial state~$\rho$  with  covariance matrix~$M(0)$.
According to~\eqref{eq:closed}, the covariance matrix $M(t)$ of the time-evolved state $\rho(t)$ satisfies
\begin{align}
\frac{d}{dt}M_{j,k}(t)=i\tr \left(c_j c_k (-i)[H,\rho]\right)=-\tr\left(\rho [H,c_jc_k]\right)\qquad \textrm{ for any }j<k\ ,
\end{align}
 where we used the cyclicity of the trace. From this expression, it follows that  the covariance matrix~$M(t)$ satisfies the equation
\begin{align}
\frac{d}{dt}M(t)=[M(t),{\bf H}]\  .\label{eq:explicitequationunitary}
\end{align}
Eq.~\eqref{eq:explicitequationunitary} has the solution  
\begin{align}
M(t)=R(t)M(0)R(t)^T\qquad\textrm{ where }\qquad R(t)=e^{-{\bf H}t}\in SO(2N,\mathbb{R})\ .\label{eq:explicitsolutionunitary}
\end{align}
The matrix $R(t)$ can be computed in time $O(N^2)$ from the Williamson normal form of~${\bf H}$. The latter can be computed in time~$O(N^3)$ (by diagonalizing ${\bf H}^T{\bf H}$), hence it follows that the evolution~$M\mapsto M(t)$ can be simulated in time~$O(N^3)$.  Eq.~\eqref{eq:explicitsolutionunitary} is sufficient for the purpose of simulating unitaries for fermionic linear optics computation starting from a Gaussian state. 

Let us discuss an alternative derivation of Eq.~\eqref{eq:explicitsolutionunitary}, which additionally provides a method for simulating the evolution of higher moments of a (possibly non-Gaussian)  initial state $\rho$. It also motivates the approach to dissipative dynamics discussed in Section~\ref{sec:thirdquantizationapproach}.   For this purpose, consider the adjoint action 
\begin{align}
c_j(t)&=e^{iHt}c_je^{-iHt}=\sum_{k=1}^{2N}R(t)_{j,k}c_k \label{eq:adjointactionhamiltonian}
\end{align}
of the unitary $e^{-iHt}$ on Majorana operators. 
Eq.~\eqref{eq:adjointactionhamiltonian}  can be shown as follows (cf.~\cite[Theorem~3]{Jozsamiyake08}). Consider the derivative
\begin{align}
\frac{dc_j(t)}{dt}=[(-i)H,c_j(t)]\ .\label{eq:cjtderiv}
\end{align}
Because $[c_{j}c_{k},c_{\ell}]=0$ unless $\ell\in\{j,k\}$ and $[c_jc_k,c_j]=-2c_k$, we have
\begin{align}
[(-i)H({\bf H}),c_\ell]&=\sum_n\frac{1}{4}{\bf H}_{m,n}[c_mc_n,c_\ell]=-\sum_n {\bf H}_{\ell,n}c_n\ .\label{eq:inserteqhcell}
\end{align}
Observe that if $c_j(t)$ is a linear combination of the operators~$\{c_\ell\}$, then so is $\frac{d}{dt}c_j(t)$ because of~\eqref{eq:cjtderiv} and~\eqref{eq:inserteqhcell}. Since this applies to $t=0$, we conclude that $c_j(t)$ is  of the
form specified on the lhs.~of~\eqref{eq:adjointactionhamiltonian} for all~$t$, that is, a linear combination of the operators $\{c_k\}$ with some coefficients~$R_{j,k}(t)$.  It remains to find the matrix~$R(t)$. Rewriting~\eqref{eq:cjtderiv} in terms of $R(t)$ and using~\eqref{eq:inserteqhcell}, we obtain
$\frac{dR(t)}{dt}=-R(t){\bf H}$
 by taking the anticommutator $\frac{1}{2}\{c_k,\cdot\}$ on both sides, for $k=1,\ldots,2N$. This shows that $-{\bf H}\in\mathfrak{so}(2N,\mathbb{R})$ indeed generates $R(t)$, proving the claim~\eqref{eq:adjointactionhamiltonian}.

Since the   covariance matrix~$M(t)$ of the time-evolved state can be computed in the Heisenberg picture as
\begin{align*}
M_{j,k}(t)&=\tr(e^{-iHt}\rho e^{iHt}c_jc_k)=\tr(\rho c_j(t)c_k(t))\ ,
\end{align*}
the claim~\eqref{eq:explicitsolutionunitary} immediately follows from~\eqref{eq:adjointactionhamiltonian}. For later reference, we point out that in this argument, we made use of~\eqref{eq:adjointactionhamiltonian} only to compute the product of two time-evolved operators, that is, in the form
\begin{align}
c_j(t)c_k(t)&=\sum_{\ell,m=1}^{2N}R(t)_{j,\ell}R(t)_{k,m}c_\ell c_m\qquad\textrm{ for }j\neq k\ .\label{eq:cjckprodcomp}
\end{align}

It is also clear how this generalizes to higher moments: for example, if 
\begin{align}
M_{j,k,\ell}(0)=\tr(\rho c_jc_kc_\ell)\label{eq:highermoment}
\end{align} are the moments of a possibly non-Gaussian state~$\rho(0)$, the moments of the time-evolved state~$\rho(t)$ are given by
\begin{align*}
M_{j,k,\ell}(t)=\sum_{j',k',\ell'}R_{j,j'}(t)R_{k,k'}(t)R_{\ell,\ell'}(t) M_{j',k',\ell'}\ .
\end{align*}

\section{Simulating dissipative evolution\label{sec:solution}}
In this section, we show how to simulate dissipative dynamics of the form~\eqref{eq:lindbladevolution}. 
 In more detail, since~\eqref{eq:lindbladevolution} preserves the Gaussian nature of a state~$\rho(0)$ according to Lemma~\ref{lem:gaussianitypreservation}, it suffices to compute the covariance matrix~$M(t)$ of the time-evolved state  $\rho(t)=e^{t\cL}(\rho)$. Our main result can be stated as follows:
\begin{theorem}\label{thm:main}
Consider $N$ fermionic modes and let $\cL$ be a Liouvillian of the form~\eqref{eq:lindbladevolution}.  Let $\rho$ be an even Gaussian state with covariance matrix~$M(0)\in\mathfrak{so}(2N,\mathbb{R})$.
The covariance matrix $M(t)$ of the time-evolved Gaussian state $\exp(t\cL)(\rho)$
can be computed in time~$O(N^3)$.
\end{theorem}
We give two different proofs of this statement: in Section~\ref{sec:evolutionequationappr}, we  sketch how to solve the corresponding differential equation directly. This approach may be most numerically stable and relies on the Bartels-Stewart algorithm for finding the covariance matrix of a fixed point. Our second method, discussed in Section~\ref{sec:thirdquantizationapproach} is based on computing the Heisenberg-evolved Majorana operators using the formalism of third quantization. This method can be adapted to simulate the evolution of higher moments starting from (possibly non-Gaussian) initial states.
\subsection{Simulation method based on evolution equation\label{sec:evolutionequationappr}}
The definition of the  covariance matrix~$M(t)$ implies that 
\be
\label{Mdot}
\frac{d}{dt} {M}_{j,k}(t)=i\, \trace{c_j c_k \cL(\rho(t))}=  i \, \trace{ \cL^\dag (c_j c_k) \rho(t)} \quad \mbox{for any $j<k$}.
\ee
The action of the adjoint Liouvillian $\cL^\dag$ (cf.~\eqref{eq:adjointliouvillian}) on observables can be written as
\[
\cL^\dag(O)=i[H,O] + \sum_\mu L_\mu^\dag [O,L_\mu] + [L_\mu^\dag,O]L_\mu.
\]
Choosing $O\sim c_j c_k$ one can easily check that the commutators $[H,O]$ and $[O,L_\mu]$ are quadratic and linear
functions of the Majorana fermion operators respectively. It follows that   $\cL^\dag$ preserves
the  subspace of operators spanned by the identity and $c_j c_k$ with $1\le j<k\le 2N$.
Therefore Eq.~(\ref{Mdot}) provides a closed linear differential equation that governs the time evolution
of $M(t)$. Simple algebra shows that
\begin{align}
\frac{d}{dt}M(t)={\bf X}M(t)+M(t){\bf X}^T+{\bf Y},\label{eq:differentialequation}
\end{align}
where
\begin{align*}
{\bf X}=-{\bf H}-2({\bf M}+{\bf M}^*)\qquad\textrm{ and }\qquad {\bf Y}=4i({\bf M}^*-{\bf M})\ .
\end{align*}
This generalizes~\eqref{eq:explicitequationunitary}.  Recall that ${\bf H}$ is a real anti-symmetric matrix of size $2N\times 2N$
parametrizing the quadratic Hamiltonian $H$, while 
${\bf M}$ is a complex Hermitian matrix of size $2N\times 2N$ parameterizing the Lindblad operators $L_\mu$,
see Section~\ref{sec:thirdquantization}. 
 It follows that ${\bf X}$ and ${\bf Y}$ are both real-valued $2N\times 2N$-matrices, and ${\bf Y}=-{\bf Y}^T$ is antisymmetric.

Since any Lindblad equation has at least one steady state\footnote{Note that for any integer $n\ge 0$
there exists at least one (mixed) state $\rho_n$ such that $e^{\cL/n}(\rho_n)=\rho_n$.
It implies $\cL(\rho_n)=O(1/n)$ for large $n$.
Since $\rho_n$ belongs to a compact manifold, the sequence $\{\rho_n\}$ has a convergent subsequence. 
Its limiting point $\rho$ is a (mixed) state that obeys $\calL(\rho)=0$.}, there must exist at least one covariance matrix $M_0$
which is a fixed point of Eq.~(\ref{eq:differentialequation}), that is,
\be
\label{Mfixed}
{\bf X}M_0+M_0{\bf X}^T+{\bf Y}=0.
\ee
Given any covariance matrix $M_0$ satisfying Eq.~(\ref{Mfixed}), the solution $M(t)$ of
Eq.~(\ref{eq:differentialequation}) can be written as
\be
\label{M(t)simple}
M(t)=M_0  + e^{{\bf X}t} (M(0)-M_0)e^{{\bf X}^Tt}, \quad t\ge 0.
\ee
Consider as an example a unitary evolution, that is, $L_\mu=0$ for all $\mu$. In this case
${\bf M}=0$, ${\bf X}=-{\bf H}$, and ${\bf Y}=0$. The fixed point covariance matrix can be chosen
as $M_0=0$ and we recover~\eqref{eq:explicitsolutionunitary}.

Finding the fixed point covariance matrix $M_0$ in the general case requires solving the system 
of linear equations Eq.~(\ref{Mfixed}), where $M_0$ is considered as an unknown vector of size
$N(2N-1)$. A na\"ive approach based on the Gaussian elimination would take time $O(N^6)$ to 
find $M_0$. However, this approach ignores the special structure of the problem. 
To the best of our knowledge, the most efficient method of solving the matrix equation
Eq.~(\ref{Mfixed}) is the Bartels-Stewart algorithm~\cite{BSmethod} which has  running time
$O(N^3)$. The key step of the algorithm is to perform a real Schur decomposition of ${\bf X}$,
that is, an orthogonal change of basis making ${\bf X}$
block upper triangular with blocks of size $1$ and $2$ on the main diagonal. 
In the new basis the resulting system of equations on matrix elements of $M_0$ turns out to be
quasi-triangular and can be solved in time~$O(N^3)$. For completeness, we sketch the Bartels-Stewart 
algorithm in Appendix~\ref{sec:BS}.

It is worth  emphasizing that all matrices involved in Eq.~(\ref{M(t)simple}) have bounded
norm independent of~$t$. This makes Eq.~(\ref{M(t)simple}) suitable for numerical calculation
of $M(t)$. Indeed, taking into account that  ${\bf X+X}^T=-4\mathrm{Re}({\bf M})\le 0$, one gets
\[
\| e^{{\bf X}t} \| \le \| e^{({\bf X +X}^T)t/2}\| \le 1.
\]
Here the first inequality follows from Theorem~IX.3.1 of~\cite{Bhatia}.

Finally, let us remark that Eq.~(\ref{eq:differentialequation}) can be solved directly
even without knowing a fixed point $M_0$ 
by transforming it into a homogeneous linear system. This method takes time only~$O(N^3)$,
but unfortunately, it is  computationally unstable for large evolution time $t$. 
Indeed, introduce an
auxiliary $2N\times 2N$ matrix $K(t)$ whose time evolution is trivial, $K(t)=K(0)=I$.
Then one can rewrite~\eqref{eq:differentialequation} as
\bea
\dot{M}(t) &=& {\bf X} M(t) + M(t) {\bf X}^T  + {\bf Y} K(t), \label{lin1} \\
\dot{K}(t) &=& -{\bf X}^T K(t) + K(t) {\bf X}^T, \label{lin2}
\eea
with  initial condition $K(0)=I$. Note that $K(t)=I$ is indeed the only  solution of
Eq.~(\ref{lin2}).  Define a  $4N\times 2N$ matrix $\Omega$ such that
\[
\Omega(t)=\left[ \ba{c} M(t) \\ K(t) \\ \ea \right].
\]
Then the system of Eqs.~(\ref{lin1},\ref{lin2}) is equivalent to
\be
\label{lin3}
\dot{\Omega}(t) = {\bf Z} \Omega(t) + \Omega(t) {\bf X}^T, \quad \mbox{where} \quad
{\bf Z}=\left[ \ba{cc} {\bf X} & {\bf Y} \\ 0 & -{\bf X}^T \\ \ea \right].
\ee
Its solution is
\be
\label{lin4}
\Omega(t)=\exp{({\bf Z}t)} \Omega(0) \exp{({\bf X}^T t)}.
\ee
This can be computed in time $O(N^3)$. Unfortunately,
using Eq.~(\ref{lin4}) for numerical calculations  may be problematic due
to the exponential growth of the factor $\exp{({\bf Z}t)}$.

\subsection{Simulation method based on `third quantization'\label{sec:thirdquantizationapproach}}
Without loss of generality, we assume that the even part~$\hat{\cL}_+=\hat{\cL}({\bf L})$ of the Liouvillian is specified by an antisymmetric matrix~${\bf L}\in\mathfrak{so}(4N,\mathbb{C})$ as in~\eqref{eq:liouvillethird}. We first derive the following analog of~\eqref{eq:adjointactionhamiltonian} (or, more precisely,~\eqref{eq:cjckprodcomp}) which applies to dissipative evolution.
\begin{lemma}\label{lem:heisenbergevolvedliouvillean}
For $j<k$, $j,k\in \{1,\ldots,2N\}$, let~$\Lambda(j,k)$ be the $4N\times 4N$~antisymmetric matrix with non-zero entries
\begin{align}
\Lambda(j,k)_{2j-1+x,2k-1+y}=i^{x+y}\qquad\textrm{ for }x,y\in\{0,1\}\ .\label{eq:omegajkdef}
\end{align}
above the main diagonal and let $\Lambda(k,j)=-\Lambda(j,k)$. Furthermore, for $p=2j-1+x$, $q=2k-1+y$, where $x,y\in\{0,1\}$, let $\Gamma(p,q)$ be the antisymmetric $2N\times 2N$-matrix with
 non-zero entries
\begin{align}
\Gamma(2j-1+x,2k-1+y)_{j,k}&=(-i)^{x+y} \qquad\textrm{ for }j<k\textrm{ and }x,y\in\{0,1\}\ \label{eq:gammadef}
\end{align}
above the main diagonal, and let $\Gamma(q,p)=-\Gamma(p,q)$.
Then
\begin{align}
\exp(t\cL_+^\dagger)(c_jc_k)&=\frac{1}{16}\sum_{r,s=1}^{4N}\sum_{\ell,m=1}^{2N}(R(t)^T\Lambda(j,k)R(t))_{r,s}\Gamma(r,s)_{\ell,m} c_{\ell}c_{m}\qquad\textrm{ for all }j<k\ .\label{eq:toproveevolve}
\end{align}
where $R(t)=\exp({\bf L}^\dagger t)$.
\end{lemma}
\begin{proof}
We have by~\eqref{eq:thirdquantizeddirac} and~\eqref{eq:thirdquantizedmajorana}
\begin{align}
\kket{c_jc_k}&=\frac{1}{8}\sum_{p,q=1}^{4N}\Lambda(j,k)_{p,q}\hat{c}_p\hat{c}_q\kket{I}\qquad\textrm{ for }j\neq k\in\{1,\ldots,2N\}\label{eq:onedirec}\\
\hat{c}_r\hat{c}_s\kket{I}&=\frac{1}{2}\sum_{\ell,m=1}^{4N}\Gamma(r,s)_{\ell,m}\kket{c_\ell c_m}\qquad\textrm{ for }r\neq s\in\{1,\ldots,4N\}\ . \label{eq:inversetransform}
\end{align}
 Using~\eqref{eq:supotimeevolve} and unitality, we get
\begin{align}
\widehat{\exp(t\cL_+^\dagger)}\hat{c}_p\hat{c}_q\kket{I}&=\hat{c}_p(t)\hat{c}_q(t)\exp(t\widehat{\cL_+^\dagger})\kket{I}=\hat{c}_p(t)\hat{c}_q(t)\kket{I}\ ,\label{eq:timeevolvedcjck}
\end{align}
where
\begin{align}
\hat{c}_p(t)&=\exp(t\widehat{\cL_+^\dagger})\hat{c}_p\exp(-t\widehat{\cL_+^\dagger})=\sum_{r=1}^{4N}R(t)_{p,r}\hat{c}_r\ \textrm{ with }\ R(t)=e^{{\bf L}^\dagger t}\in SO(4N,\mathbb{C})\ \label{eq:heisenbergchat}
\end{align}
is the Heisenberg time-evolved super-Majorana operator $\hat{c}_p$. Equation~\eqref{eq:heisenbergchat}
follows
in the same way as~\eqref{eq:adjointactionhamiltonian} because we have
\begin{align}
[\cL({\bf L}),\hat{c}_{\ell}]=\sum_{k}{\bf L}_{\ell,k}\hat{c}_k\ .
\end{align}
in analogy with~\eqref{eq:inserteqhcell}.  Reinserting~\eqref{eq:heisenbergchat} into~\eqref{eq:timeevolvedcjck} gives
\begin{align}
\widehat{\exp(t\cL_+^\dagger)}\hat{c}_p\hat{c}_q\kket{I}&=\sum_{r\neq s}^{4N} R_{p,r}(t)R_{q,s}(t)\hat{c}_r \hat{c}_s\kket{I}\qquad\textrm{ for }p\neq q\ .\label{eq:timeevolvedcpcqthird}
\end{align}
Here we can restrict the sum to $r\neq s$ because~$R(t)$ is orthogonal and $\hat{c}_r^2=I$ for all~$r$.
Equation~\eqref{eq:timeevolvedcpcqthird} is formally analogous to~\eqref{eq:cjckprodcomp}.  The claim follows by computing $\widehat{\exp(t\cL_+^\dagger)}\kket{c_jc_k}$ using expression~\eqref{eq:onedirec}, linearity and~\eqref{eq:timeevolvedcpcqthird}, and then translating the result back with~\eqref{eq:inversetransform}.
 \end{proof}

We can use Lemma~\ref{lem:heisenbergevolvedliouvillean} to compute  the time-evolved covariance matrix~$M(t)$ of a Gaussian state~$\rho$ with covariance matrix~$M(0)$. Expressing the matrix elements using the Heisenberg picture gives
\begin{align}
M_{j,k}(t)=i\tr(\rho \exp(t\cL_+^\dagger)(c_jc_k))\qquad\textrm{ for }j<k\ \label{eq:mjkmatrixelement}
\end{align}
since $\exp(t\cL)(\rho)=\exp(t\cL_+)(\rho)$. Theorem~\ref{thm:main}
essentially follows by combining this equation with Lemma~\ref{lem:heisenbergevolvedliouvillean}. However, showing that the resulting expressions can be evaluated in time~$O(N^3)$ requires some care.

 Let $R=R(t)=\exp({\bf L}^\dagger t)$ be as in Lemma~\ref{lem:heisenbergevolvedliouvillean} and let $\Lambda(j,k)$ and~$\Gamma(r,s)$
 be defined by~\eqref{eq:omegajkdef} and~\eqref{eq:gammadef}, respectively.
 With Lemma~\ref{lem:heisenbergevolvedliouvillean} and~\eqref{eq:mjkmatrixelement} we have
\begin{align}
M_{j,k}(t)=\frac{1}{16}\sum_{r,s=1}^{4N}\sum_{\ell,m=1}^{2N} (R^T\Lambda(j,k)R)_{r,s} \Gamma(r,s)_{\ell,m}M_{\ell,m}(0)=\frac{1}{16} \tr\left((R^T\Lambda(j,k)R) \Omega(0)^T\right)\ ,\label{eq:mjkmatrixelementcomp}
\end{align}
where $\Omega(0)$ is the $4N\times 4N$-matrix defined by the entries
\begin{align}
\Omega_{r,s}(0)=\tr\left(\Gamma(r,s)M(0)^T\right)\ .\label{eq:bmatrixdef}
\end{align}
Using the general identity $\tr\left((ABC)D^T\right)=\tr\left((A^TDC^T)B^T\right)$, we can rewrite~\eqref{eq:mjkmatrixelementcomp} as 
$M_{j,k}(t)=\frac{1}{16}\tr\left((R \Omega(0) R^T)\Lambda(j,k)^T\right)$ 
or
\begin{align}
M_{j,k}(t)=\frac{1}{16}\tr\left(\Omega(t)\Lambda(j,k)^T\right)\label{eq:mjkabexpromega}
\end{align}
where
\begin{align}
\Omega(t)=\exp({\bf L}^\dagger t)\Omega(0) \exp({\bf L}^\dagger t)^T.\label{eq:omegatdef}
\end{align}
Clearly, $\Omega(0)$ can be obtained from $M(0)$ in time $O(N^2)$ since every entry~$\Omega_{r,s}(0)$  is a linear combination of a constant number of entries of~$M$ (cf.~\eqref{eq:bmatrixdef}). Similarly, the matrix~$M(t)$ can be computed in time $O(N^2)$ in an entrywise fashion according to~\eqref{eq:mjkabexpromega} given the matrix~$\Omega(t)$. Therefore, the claim of Theorem~\ref{thm:main} follows since~$\Omega(t)$ can be computed from~$\Omega(0)$  in time $O(N^3)$ using~\eqref{eq:omegatdef}.  To avoid exponentially growing terms arising from the expression $\exp({\bf L}^\dagger t)$ in~\eqref{eq:omegatdef}, we can decompose $\Omega(t)$ as
\begin{align}
\Omega(t)=\Omega_{b}(t)+\Omega_{u}(t)\ \label{eq:jordanboundedunbounded}
\end{align}
using the Jordan decomposition of ${\bf L}^\dagger$, such that the matrix elements of $\Omega_{b}(t)$ are bounded for all $t$, whereas the matrix elements of $\Omega_u(t)$ are exponentially growing with~$t$. Because $M(t)$  is the covariance matrix of the time-evolved state~$\exp(t\cL)(\rho)$, its entries are bounded for all~$t$ and we conclude from~\eqref{eq:mjkabexpromega} that the contribution of $\Omega_u(t)$ must vanish, i.e.,  
\begin{align*}
M_{j,k}(t)=\frac{1}{16}\tr\left(\Omega_b(t)\Lambda(j,k)^T\right)\ .
\end{align*}
To describe the decomposition~\eqref{eq:jordanboundedunbounded} in more detail, 
consider
the Jordan normal form of ${\bf L}^\dagger$,
\begin{align}
{\bf L}^\dagger&= V \bigoplus_{\alpha}J_{\alpha}(\lambda_\alpha) V^{-1}\ ,\label{eq:jordannormalform}
\end{align}
where $V$ is an invertible matrix. Here $J_\alpha(\lambda_\alpha)$ is a Jordan block
\begin{align*}
J_{\alpha}(\lambda_\alpha)&= \lambda_\alpha I_{m_\alpha}+N_{m_\alpha}
\end{align*}
where $I_{m_\alpha}$ is the identity matrix of size~$m_\alpha\times m_\alpha$  and $N_{m_\alpha}$ is the nilpotent matrix with ones on the first upper off-diagonal. With
\begin{align*}
\exp(J_\alpha(\lambda_\alpha) t)&=e^{\lambda_\alpha t}\left(I_{m_\alpha}+\sum_{k=1}^{m_\alpha-1} \frac{t^k}{k!}N_{m_\alpha}^k\right)\equiv e^{\lambda_\alpha t} S_{m_\alpha}(t)\ ,
\end{align*}
and $\exp(-J_\alpha(\lambda_\alpha)t)=e^{-\lambda_\alpha t}S_{m_\alpha}(-t)$ we get
\begin{align}
\Omega(t)&=V\left(\bigoplus_{\alpha,\beta} e^{(\lambda_\alpha-\lambda_\beta)t}S_{m_\alpha}(t) V^{-1}\Omega(0)V S_{m_\beta}(-t)\right) V^{-1}\ .\label{eq:rbrtrexpr}
\end{align}
Clearly, all terms corresponding to pairs $(\alpha,\beta)$ with $Re(\lambda_\alpha-\lambda_\beta)>0$ grow exponentially and must be absorbed in~$\Omega_u(t)$. Furthermore, for pairs with $Re(\lambda_\alpha-\lambda_\beta)=0$, the contribution of every nilpotent term~$N^k_m$ grows polynomially with~$t$. The remaining terms remain bounded, and we conclude that $\Omega_b(t)$ is given by
\begin{align}
\Omega_b(t) = &V\left(\bigoplus_{\alpha,\beta: Re(\lambda_\alpha-\lambda_\beta)<0} e^{(\lambda_\alpha-\lambda_\beta)t}S_{m_\alpha}(t) V^{-1}\Omega(0)V S_{m_\beta}(-t)\right) V^{-1}\nonumber\\
+&V\left(\bigoplus_{\alpha,\beta: Re(\lambda_\alpha-\lambda_\beta)=0} e^{(\lambda_\alpha-\lambda_\beta)t}I_{m_\alpha} V^{-1}\Omega(0)V I_{m_\beta}\right) V^{-1}\ .\label{eq:omegabexpression}
\end{align}
Because computing the Jordan normal form~\eqref{eq:jordannormalform}  takes time~$O(N^3)$, the matrix~$\Omega_b(t)$ (and hence also $M(t)$) can be computed in time~$O(N^3)$ using~\eqref{eq:omegabexpression}.

Finally, let us sketch the modifications required to simulate 
the evolution of higher moments such as~\eqref{eq:highermoment} starting from a non-Gaussian initial state. Clearly, Lemma~\ref{lem:heisenbergevolvedliouvillean} can be generalized to give an expansion of the Heisenberg evolved product $\exp(t\cL^\dagger_+)(c_jc_kc_\ell)$ in terms of a linear combination of Majorana monomials~$c_{j'}c_{k'}c_{\ell'}$.  Inserting this into 
$M_{j,k,\ell}(t)=\tr(\rho \exp(t\cL^\dagger_+)(c_jc_kc_\ell))$  immediately gives an explicit expression for the tensor of time-evolved moments $M(t)$ in terms of the original moments~$M(0)$.

\appendix

\section{Computing the steady state\label{sec:BS}}
In this appendix we sketch the Bartels-Stewart algorithm~\cite{BSmethod} for solving a matrix equation
\be
\label{fixed}
{\bf X}M+M{\bf X}^T+{\bf Y}=0
\ee
which determines the steady state covariance matrix $M\equiv M_0$, see Eq.~\eqref{Mfixed} in Section~\ref{sec:evolutionequationappr}.
Here ${\bf X}$ and ${\bf Y}$ are known real matrices such that ${\bf Y}^T=-{\bf Y}$, while $M$
is an unknown matrix to be found. All matrices have size $n\times n$.
As opposed to the third quantization method,
where computing the steady state may involve manipulations with ill-conditioned invertible matrices, see Eq.~(\ref{eq:jordannormalform}), the Bartels-Stewart method uses only transformations based on orthogonal (unitary) matrices.
This is likely to offer a better numerical stability.

The first step of the algorithm involves the Schur decomposition\footnote{To simplify notations, we slightly
 deviate from the original algorithm of~\cite{BSmethod} which adopted a real Schur decomposition.} of ${\bf X}$. This defines a unitary
matrix $U$ such that $X=U^\dagger{\bf X}U$ is upper triangular.
Introducing a new unknown matrix $K=U^\dag M U$ we can
rewrite Eq.~(\ref{fixed}) as
\be
\label{fixed1}
XK+KX^\dag =Y
\ee
where $Y=-U^\dag {\bf Y} U$.
Given any matrix $Z$, the $j$-th column of $Z$ will be denoted $Z_j$. Taking the $n$-th column of Eq.~(\ref{fixed1})
one arrives at
\be
\label{fixed2}
(X+X_{n,n}^* \, I ) K_n = Y_n.
\ee
This is a triangular linear system of equations for the  unknown vector $K_n$ which can be solved
in time $O(n^2)$. Taking the $m$-th column of Eq.~(\ref{fixed1}) one arrives at
\be
\label{fixed3}
(X+X_{m,m}^*\, I) K_m = Y_m - \sum_{j=m+1}^n X_{m,j}^* K_j, \quad m=n-1,\ldots,2,1.
\ee
Suppose the columns $K_{m+1},\ldots,K_n$ are already known. Then the right-hand side
of Eq.~(\ref{fixed3}) can be formed in time $O(n^2)$. We get a triangular system of equations 
for the unknown vector $K_m$ which can be solved in time $O(n^2)$. 
Proceeding inductively from $m=n-1$ towards $m=1$ we can compute the entire matrix $K$
in time $O(n^3)$. This gives a solution of Eq.~(\ref{fixed}), namely $M=U K U^\dag$.
In general, this solution is neither real nor anti-symmetric. However, one can easily check that
matrices $(M+M^*)/2$ and $(M-M^T)/2$ are solutions of Eq.~(\ref{fixed})
for any (complex) solution $M$.
Hence we can  always transform $M$ into a real anti-symmetric solution. 

Computing the Schur decomposition of a matrix ${\bf X}$ is a standard subroutine available in 
many numerical linear algebra tools (such as MATLAB). Theoretically, it can be computed in time
$O(n^3)$ by finding generalized eigenvectors of ${\bf X}$ and applying the Gram-Schmidt orthogonalization.
A more practical method used in~\cite{BSmethod} involves two steps. First, one 
transforms ${\bf X}$ to the upper Hessenberg form by a sequence of $n-2$  Householder reflections,
see~\cite[p.~346]{Wilkinson}, which requires $O(n^3)$ elementary  operations.  
Secondly, ${\bf X}$ is made upper triangular by the QR-algorithm. Each iteration of the QR-algorithm
requires $O(n^2)$ elementary operations, however the required number of iterations is generally unknown.

\subsection*{Acknowledgments}
The authors acknowledge partial support by the DARPA QuEST program under contract number~HR0011-09-C-0047.

\bibliography{q}

\end{document}